\pdfoutput=1
\documentclass[journal]{IEEEtran}

\IEEEoverridecommandlockouts
\usepackage{cite}
\usepackage{amsmath,amssymb,amsfonts}
\usepackage{algorithmic}
\usepackage{graphicx}
\usepackage{textcomp}
\usepackage{xcolor}
\usepackage{float}
\usepackage{booktabs}
\usepackage{subcaption}
\usepackage{algorithm}

\def\BibTeX{{\rm B\kern-.05em{\sc i\kern-.025em b}\kern-.08em
    T\kern-.1667em\lower.7ex\hbox{E}\kern-.125emX}}

\begin{document}

\title{Reducing Complexity for Quantum Approaches in Train Load Optimization}

\author{
\IEEEauthorblockN{Zhijie Tang\IEEEauthorrefmark{1}}, 
\IEEEauthorblockN{Albert Nieto-Morales\IEEEauthorrefmark{2}}, and 
\IEEEauthorblockN{Arit Kumar Bishwas\IEEEauthorrefmark{3}\textsuperscript{\textdagger}} \\
\IEEEauthorblockA{PricewaterhouseCoopers} \\
\IEEEauthorblockA{\IEEEauthorrefmark{1}Commercial Technology and Innovation, Shanghai, China (e-mail: zhijietang0710@163.com)} \\
\IEEEauthorblockA{\IEEEauthorrefmark{2}Tech and Innovation, San Francisco, US (e-mail: albert.morales@pwc.com)} \\
\IEEEauthorblockA{\IEEEauthorrefmark{3}Tech and Innovation, San Francisco, US (e-mail: arit.kumar.bishwas@pwc.com)}
\thanks{\textdagger~Corresponding author.}
}

\maketitle

\begin{abstract}
Efficiently planning container loads onto trains is a computationally challenging combinatorial optimization problem, central to logistics and supply chain management. A primary source of this complexity arises from the need to model and reduce rehandle operations—unproductive crane moves required to access blocked containers. Conventional mathematical formulations address this by introducing explicit binary variables and a web of logical constraints for each potential rehandle, resulting in large-scale models that are difficult to solve. This paper presents a fundamental departure from this paradigm. We introduce an innovative and compact mathematical formulation for the Train Load Optimization (TLO) problem where the rehandle cost is calculated implicitly within the objective function. This novel approach helps prevent the need for dedicated rehandle variables and their associated constraints, leading to a dramatic reduction in model size. We provide a formal comparison against a conventional model to analytically demonstrate the significant reduction in the number of variables and constraints. The efficacy of our compact formulation is assessed through a simulated annealing metaheuristic, which finds high-quality loading plans for various problem instances. The results confirm that our model is not only more parsimonious but also practically effective, offering a scalable and powerful tool for modern rail logistics.
\end{abstract}


\IEEEpeerreviewmaketitle

\section{Introduction}
\IEEEPARstart{T}{he} relentless growth of global trade has solidified the role of containerized cargo as the backbone of international logistics \cite{Rodrigue2020}. Within the multimodal transport network, rail plays an important role due to its high capacity, energy efficiency, and reliability for inland transportation \cite{Woodburn2006, Janic2007, Facchini2016}. Rail networks offer strong intermodal connectivity with ports and trucking, forming an integrated system for global supply chains \cite{Bierwirth2010, Macharis2008}. To harness these advantages, operators must solve the Train Load Optimization (TLO) problem, which involves assigning containers from a yard to slots on a train to maximize value while respecting a multitude of operational and physical constraints \cite{Bostel2007}. Effective TLO leads to increased asset utilization, reduced costs, and improved service reliability \cite{Kozan2000}.

One of the more significant operational inefficiencies in container terminals is the rehandle (or reshuffle) operation \cite{Zhang2002}. Containers are typically stored in vertical stacks in the yard \cite{Stahlbock2008, Carlo2014}; if a container selected for loading is not at the top of its stack, all containers above it must be moved. If these blocking containers are also not destined for the current loading operation, they should be temporarily placed elsewhere, incurring costs in terms of time, fuel, and crane utilization \cite{Dekker2006, Lee2009}. Minimizing these rehandles is one of the primary objectives in TLO and related yard management problems \cite{Goodchild2005}.

The academic literature has extensively studied TLO and related terminal operations problems \cite{Boysen2017, Bierwirth2015}. A common strategy for modeling rehandles in integer programming formulations is to introduce explicit binary variables that are activated when a rehandle occurs \cite{Vacca2013, Caserta2011}. This approach requires a detailed position matrix to track the vertical relationships between containers and a set of logical "Big-M" constraints to enforce the rehandle logic. While this method is mathematically sound, it suffers from a major drawback: scalability. As the number of containers and wagons increases, the number of rehandle variables and constraints grow polynomially, leading to massive models that overwhelm even industry-leading commercial solvers, a common issue in large-scale logistics problems \cite{Crainic2009}.

This paper challenges the conventional modeling paradigm. We propose a novel mathematical formulation for the TLO problem that fundamentally rethinks how rehandles are calculated. Our main contribution is the development of a compact model that calculates the total rehandle cost \emph{implicitly} within the objective function itself. By leveraging the sequential nature of train loading, our formulation directly counts the number of necessary rehandles based on the assignment decisions, without needing any auxiliary rehandle variables or constraints. This innovation leads to a provably simpler and smaller model. We formally present both the conventional and our proposed formulations to highlight the structural differences. We then employ a simulated annealing metaheuristic \cite{Kirkpatrick1983, Cerny1985}, an innovative tool for complex combinatorial problems \cite{Glover1989}, to demonstrate that our more compact model can be solved effectively to produce high-quality, practical loading plans. The results of our experiments assess the core thesis of our work: by simplifying the mathematical representation of the problem, we can create more scalable and effective tools for real-world TLO challenges.

\begin{figure*}[!t]
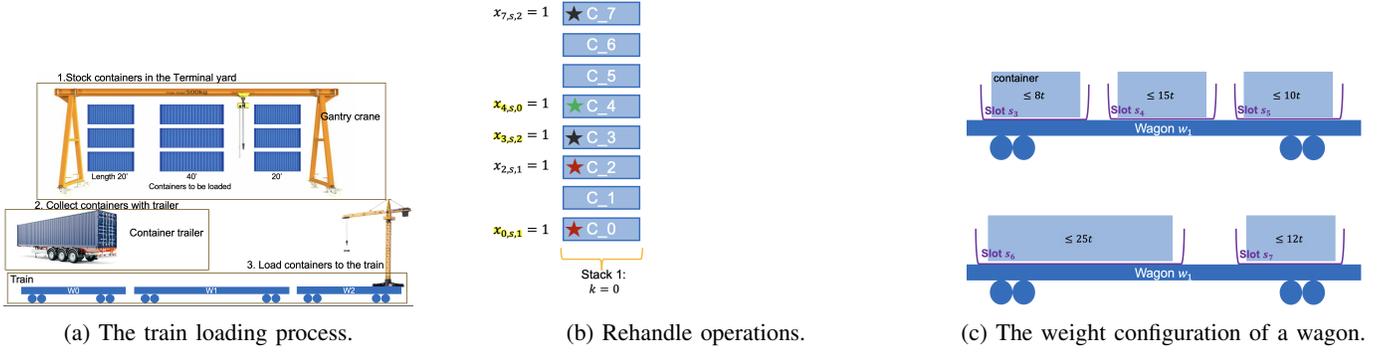

\centering
\begin{subfigure}[t]{0.3\textwidth}
    \includegraphics[width=\linewidth]{train_load_process.pdf}
    \caption{The train loading process.}
    \label{fig:loading_process}
\end{subfigure}\hfill
\begin{subfigure}[t]{0.3\textwidth}
    \includegraphics[width=0.45\linewidth]{rehandle_operation.pdf}
    \caption{Rehandle operations.}
    \label{fig:rehandle_op}
\end{subfigure}\hfill
\begin{subfigure}[t]{0.3\textwidth}
    \includegraphics[width=\linewidth]{weight_config.pdf}
    \caption{The weight configuration of a wagon.}
    \label{fig:weight_config}
\end{subfigure}
\caption{An overview of key operations in the train loading process, from yard stacking to container rehandling and wagon configuration.}
\label{fig:operations_overview}
\end{figure*}

\section{Problem Definition}
The Train Load Optimization problem addressed in this paper involves assigning a subset of containers from a storage yard to available slots on a train. The yard consists of multiple stacks, each containing containers tiered vertically \cite{Stahlbock2008}. Containers are moved from the yard to the rail park via internal vehicles \cite{Guan2002}. The train consists of multiple wagons, each with a specific set of slots that can accommodate either 20-foot or 40-foot containers.

The optimization objective is twofold: to maximize the total value of the loaded containers (or equivalently, minimize the penalty for containers not loaded) and to minimize the total cost of rehandle operations incurred during loading \cite{Kim2004}. This can be achieved subject to several constraints:
\begin{enumerate}
    \item Each container can be assigned to at most one slot, and each slot can hold at most one container.
    \item The total weight of containers assigned to each wagon must not exceed the wagon's capacity. Furthermore, the total weight of all containers on the train must not exceed the train's overall capacity, respecting axle load limits \cite{Ayala-Quinones2020}.
    \item Each wagon must be assigned a specific weight configuration, which dictates the maximum permissible weight for each of its individual slots (as shown in Fig.~\ref{fig:weight_config}).
    \item A rehandle is required for any container that is positioned above a target container in the same stack, if that blocking container is not itself being loaded onto the same or an earlier wagon.
\end{enumerate}

We consider a common operational scenario where wagons are serviced sequentially by a crane, from the first to the last. The parameters and sets used to model this problem are formally defined in Table \ref{tab:notation}. For the conventional model, we add a parameter $P_{ij}$ derived from the yard layout, where $P_{ij}=1$ if container $j$ is stacked above container $i$.

\begin{table}[h]
\centering
\caption{Notation for Model Parameters and Sets}
\label{tab:notation}
\begin{tabular}{@{}ll@{}}
\toprule
\textbf{Symbol} & \textbf{Description} \\ \midrule
\multicolumn{2}{l}{\textbf{Sets and Indices}} \\
$\mathcal{C}$ & Set of containers, indexed by $i, j$ \\
$\mathcal{W}$ & Set of wagons, indexed by $w, h$ \\
$S_w$ & Set of slots for wagon $w$, indexed by $s$ \\
$B_w$ & Set of weight configurations for wagon $w$, indexed by $b$ \\
$k$ & Index for container stacks, $k = 1, \ldots, K$ \\
\multicolumn{2}{l}{\textbf{Parameters}} \\
$P_{ij}$ & 1 if container $j$ is above $i$ in the same stack \\
$\omega_i$ & Weight of container $i \in \mathcal{C}$ \\
$\pi_i$ & Value (penalty for not loading) of container $i \in \mathcal{C}$ \\
$\alpha$ & Unitary cost for a rehandle operation \\
$T$ & Maximum number of tiers per stack \\
$\Omega_w$ & Maximum weight capacity of wagon $w \in \mathcal{W}$ \\
$\Omega$ & Maximum weight capacity of the entire train \\
$\delta_{b,s}$ & Max weight for slot $s$ under configuration $b$ \\
$\lambda_i$ & Length of container $i$ (e.g., 20' or 40') \\
$\mu_{s,w}$ & Length of slot $s$ in wagon $w$ \\ \bottomrule
\end{tabular}
\end{table}

\section{Mathematical Modeling Approaches}
In this section, we formally present two distinct mathematical formulations for the TLO problem. The first represents a conventional approach found in the literature, while the second is our proposed compact formulation.

\subsection{A Conventional Formulation with Explicit Rehandles (Model A)}
This standard modeling approach relies on three sets of binary decision variables to capture the state of the system:
\begin{itemize}
    \item $x_{i,s,w} \in \{0, 1\}$: 1 if container $i$ is assigned to slot $s$ of wagon $w$.
    \item $t_{w,b} \in \{0, 1\}$: 1 if weight configuration $b$ is chosen for wagon $w$.
    \item $y_{i,w} \in \{0, 1\}$: 1 if container $i$ must be rehandled to access a container below it during the loading phase of wagon $w$.
\end{itemize}
The objective function seeks to minimize the total cost, defined as the sum of rehandle costs and the penalty for unloaded containers:
\begin{equation}
\min \> \alpha \sum_{i \in \mathcal{C}} \sum_{w \in \mathcal{W}} y_{i,w} + \sum_{i \in \mathcal{C}} \pi_i \left(1 - \sum_{s,w} x_{i,s,w}\right)
\end{equation}
The model is subject to a series of constraints. Basic assignment and capacity constraints are given by (\ref{eq:assign_container})--(\ref{eq:train_weight_constraint}), which will be detailed in the next subsection. The core of this model's complexity lies in the rehandle logic constraint:
\begin{align}
\sum_{s \in S_w} x_{j,s,w} \le T \cdot \left(y_{i,w} + \sum_{h \le w} \sum_{s \in S_h} x_{i,s,h}\right), \nonumber \\ 
\forall w \in \mathcal{W}, \forall (i,j) \text{ s.t. } P_{ij}=1 \label{eq:rehandle_constraint}
\end{align}
This "Big-M" constraint is the linchpin that connects the assignment decisions to the rehandle variables. It states that if a container $j$ (which is above $i$) is assigned to wagon $w$, then the term on the right-hand side must be at least 1. This can only be satisfied if either the blocked container $i$ has already been loaded by wagon $w$ or an earlier one (making the second summation 1), or the rehandle variable $y_{i,w}$ is forced to be 1. The constant $T$ (maximum tiers) serves as a sufficiently large number. This formulation requires one such constraint for every pair of containers where one is above the other, leading to a rapid proliferation of constraints.

\subsection{The Proposed Compact Formulation with Implicit Rehandles (Model B)}
Our proposed model achieves a significant simplification by eliminating the $y_{i,w}$ variables entirely. The decision variables are reduced to:
\begin{itemize}
    \item $x_{i,s,w} \in \{0, 1\}$: 1 if container $i$ is assigned to slot $s$ of wagon $w$.
    \item $t_{w,b} \in \{0, 1\}$: 1 if weight configuration $b$ is chosen for wagon $w$.
\end{itemize}

The innovation lies in a reformulated objective function that directly computes the rehandle cost from the assignment variables $x_{i,s,w}$. We assume an indexing scheme where container $i = T \cdot k + l$ refers to the container in stack $k$ at tier $l$ (where $l=0$ is the bottom tier). The objective function is:
\begin{align}
\min \> & \sum_{k=0}^{K-1} \sum_{w=0}^{W-1} \sum_{l=0}^{T-1} \alpha \cdot L_{l,k,w} \cdot \label{eq:objective_function} \\
& \left( (T - 1 - l) - \sum_{a=l+1}^{T-1} \sum_{h=0}^{w} \sum_{s \in S_h} x_{T \cdot k + a, s, h} \right) \nonumber \\
& + \sum_{i \in \mathcal{C}} \pi_i \left(1 - \sum_{s,w} x_{i,s,w}\right) \nonumber
\end{align}
Here, $L_{l,k,w}$ is an auxiliary binary term, formally defined as $L_{l,k,w} = (\sum_{s} x_{T \cdot k+l, s, w}) \cdot (1 - \sum_{h<w}\sum_{s} x_{T \cdot k+l, s, h})$. It is 1 only if the container at $(k,l)$ is assigned for the first time at the wagon $w$ stage. The term $(T-1-l)$ is the initial number of containers blocking it. The triple summation subtracts from this count every blocking container that is also loaded by wagon $w$ or any preceding wagon. The result is the exact number of rehandles needed to retrieve the container at $(k,l)$ for wagon $w$. This formulation elegantly captures the rehandle logic without auxiliary variables or constraints.

The objective function is subject to the following set of standard operational constraints:
\begin{align}
    &\sum_{w \in \mathcal{W}} \sum_{s \in S_w} x_{i,s,w} \leq 1, \quad \forall i \in \mathcal{C} \label{eq:assign_container}\\
    &\sum_{i \in \mathcal{C}} x_{i,s,w} \leq 1, \quad \forall w \in \mathcal{W}, \, \forall s \in S_w \label{eq:one_container_per_slot}\\    
    &\sum_{b \in B_w} t_{w,b} = 1, \quad \forall w \in \mathcal{W} \label{eq:one_configuration_per_wagon}\\
    &\sum_{i \in \mathcal{C}} \omega_i x_{i,s,w} \leq \sum_{b \in B_{w}} \delta_{b,s} t_{w,b}, \quad \forall w, s \label{eq:slot_weight_constraint}\\
    &\sum_{i \in \mathcal{C}} \sum_{s \in S_w} \omega_i x_{i,s,w} \leq \Omega_w, \quad \forall w \in \mathcal{W} \label{eq:wagon_weight_constraint}\\
    &\sum_{i \in \mathcal{C}} \sum_{w \in \mathcal{W}} \sum_{s \in S_w} \omega_i x_{i,s,w} \leq \Omega \label{eq:train_weight_constraint}
\end{align}
Constraint~(\ref{eq:assign_container}) ensures each container is assigned at most once. (\ref{eq:one_container_per_slot}) allows at most one container per slot. (\ref{eq:one_configuration_per_wagon}) selects exactly one weight configuration per wagon. Constraints (\ref{eq:slot_weight_constraint})–(\ref{eq:train_weight_constraint}) enforce the weight limits for slots, wagons, and the overall train.

\section{Simulated Annealing Solution Method}
Given the NP-hard nature of the TLO problem, metaheuristics provide a practical approach for finding high-quality solutions in a reasonable amount of time. We employ Simulated Annealing (SA), a well-established, trajectory-based metaheuristic that effectively explores complex solution spaces \cite{Aarts1989, Suman2006}. SA iteratively transitions from a current solution to a randomly generated neighboring solution. Moves that improve the objective function are always accepted. Critically, moves that worsen the objective function are accepted with a probability governed by the Boltzmann distribution, $P(\text{accept}) = e^{-\Delta E / T}$, where $\Delta E$ is the cost increase and $T$ is a control parameter called temperature \cite{Dowsland2007}.

This probabilistic acceptance allows the algorithm to escape local optima, especially at high initial temperatures. As the algorithm progresses, $T$ is gradually lowered according to a cooling schedule, reducing the probability of accepting bad moves and allowing the search to converge on a promising region. The high-level process is outlined in Algorithm 1.

It is important to note that our SA implementation explores the constrained assignment and configuration solution space via composite neighborhood moves, distinguishing it from QUBO-based simulated annealing which would require an unconstrained bit-vector representation and penalty terms.

For our implementation, the neighborhood of a solution is defined by three types of moves:

\begin{enumerate} 
\item Swap move, where the assignments of two randomly selected containers are swapped. This can be a swap between two assigned containers, or between an assigned and an unassigned container.
\item Relocation move, where a randomly selected assigned container is moved to a different, randomly selected, valid, and empty slot.
\item Configuration move, where for a randomly selected wagon, its current weight configuration is changed to another valid configuration from its set of options. 
\end{enumerate}

At each step, one of these moves is chosen randomly to generate a neighbor solution. The SA parameters, including an initial temperature of 1000, a cooling rate of 0.95, and 100 iterations per temperature level, were set based on preliminary computational tests to ensure a good balance between exploration and exploitation.

\begin{algorithm}
\caption{Simulated Annealing for TLO}
\begin{algorithmic}[1]
\STATE $s_{current} \leftarrow \text{GenerateInitialSolution()}$
\STATE $s_{best} \leftarrow s_{current}$
\STATE $T \leftarrow T_{initial}$
\WHILE{$T > T_{final}$}
    \FOR{$i = 1$ to $N_{iterations}$}
        \STATE $s_{new} \leftarrow \text{GenerateNeighbor}(s_{current})$
        \STATE $\Delta E \leftarrow \text{Objective}(s_{new}) - \text{Objective}(s_{current})$
        \IF{$\Delta E < 0$ or $\text{random}(0, 1) < e^{-\Delta E / T}$}
            \STATE $s_{current} \leftarrow s_{new}$
        \ENDIF
        \IF{$\text{Objective}(s_{current}) < \text{Objective}(s_{best})$}
            \STATE $s_{best} \leftarrow s_{current}$
        \ENDIF
    \ENDFOR
    \STATE $T \leftarrow T \times \gamma$ \COMMENT{Cooling schedule}
\ENDWHILE
\RETURN $s_{best}$
\end{algorithmic}
\end{algorithm}

\section{Experimental Analysis}
Our experimental study was conducted on a PC with a 2.3 GHz Quad-Core Intel Core i7 processor. We created three test instances with varying scales to represent small, medium, and large operational scenarios, as described in Table \ref{Instances}. The analysis aims to first analytically quantify the complexity reduction achieved by our proposed model, and second, to empirically validate its performance when solved with the SA algorithm.

\begin{table}
\centering
\caption{Instances characteristics}
\label{Instances}
\begin{tabular}{@{}cccccc@{}}
    \toprule
    \textbf{Instance} & \textbf{\# cont.} & \textbf{\# wagons} & \textbf{\# tiers} & \textbf{TEUs (Train)} & \textbf{Total TEUs} \\
    \midrule
1 & 6 & 1 & 3 & 2 & 9 \\
2 & 12 & 2 & 4 & 5 & 18 \\
3 & 20 & 8 & 4 & 19 & 28 \\
\bottomrule
\end{tabular}
\end{table}

\subsection{Comparison of Model Complexity}
The primary theoretical contribution of our work is the development of a more parsimonious model. Table \ref{Comparison} provides a stark comparison of the sizes of the conventional formulation (Model A) and our proposed compact formulation (Model B) for the largest test instance. The number of variables in Model A includes the assignment variables ($x_{i,s,w}$), configuration variables ($t_{w,b}$), and rehandle variables ($y_{i,w}$). The number of constraints includes the basic constraints plus one rehandle constraint for each blocking container pair. Model B eliminates the $y_{i,w}$ variables and all rehandle constraints.

The results are compelling: our proposed model reduces the number of variables by over 50\% and the number of constraints by over 80\%. This is not merely an incremental improvement; it is a structural simplification that fundamentally reduces the dimensionality of the problem's mathematical representation. This reduction is expected to make the problem significantly more tractable for any solution method, whether exact or heuristic.

\begin{table}
\centering
\caption{Model Complexity Comparison for Instance 3}
\label{Comparison}
\begin{tabular}{@{}lcc@{}}
    \toprule
    \textbf{Model} & \textbf{\# variables} & \textbf{\# constraints}  \\
    \midrule
A (Conventional)   & $\sim$760 & $\sim$580  \\
B (Proposed)   & 360 & 97  \\
\bottomrule
\end{tabular}
\end{table}

\subsection{Performance of the Proposed Model}
Having established the theoretical compactness of Model B, we now assess its practical performance. Table \ref{Results} summarizes the results obtained by applying our SA algorithm to solve Model B for the three test instances. The metrics reported include the final objective value, the number of rehandles, slot utilization (L\%), TEU utilization ($\bar{L}$\%), the percentage of total container value loaded (P\%), and the computation time in seconds.

For the small and medium instances, the algorithm consistently found optimal or near-optimal solutions with zero rehandles and 100\% slot utilization in a matter of seconds. For the more challenging large-scale instance (Instance 3), the algorithm converged to a high-quality solution in approximately 10 minutes. This solution achieved 100\% slot utilization for the loaded wagons and incurred only 6 rehandles, loading nearly 67\% of the total available TEUs and cargo value. The runtime demonstrates manageable growth, indicating that the approach remains practical even as problem size increases. These results strongly suggest that our compact formulation, paired with an effective metaheuristic, is a viable and powerful approach for solving real-world TLO problems.

\begin{table}[H]
\centering
\caption{Results for Proposed Model B using Simulated Annealing}
\label{Results}
\begingroup
\small
\setlength{\tabcolsep}{3pt}
\begin{tabular}{@{}ccccccccc@{}}
\toprule
\textbf{Inst.} & \textbf{Vars} & \textbf{Constr.} & \textbf{Obj.} & \textbf{Reh.} & \textbf{L(\%)} & \textbf{$\bar{L}$(\%)} & \textbf{P(\%)} & \textbf{Time (s)} \\
\midrule
1 & 13 & 15 & $-13$ & 0 & 100 & 22.22 & 20 & 0.51\\
2 & 57 & 33  & $-40$ & 0 & 100 & 27.78 & 28.37 & 6.36\\
3 & 360 & 97  & $-125$ & 6 & 100 & 67.86 & 66.81 & 650.79\\
\bottomrule
\end{tabular}
\endgroup
\end{table}

\section{Prospects for Quantum Annealing Optimization}

The structural simplification of the Proposed Compact Formulation (Model B) is a critical enabler for exploring the TLO problem with Quantum Annealing (QA). QA is a specialized optimization technique requiring the problem to be mapped onto a Quadratic Unconstrained Binary Optimization (QUBO) model. Recently, QA has demonstrated significant potential in efficiently solving diverse and complex combinatorial optimization problems across various domains, ranging from portfolio optimization \cite{tang2024comparativeanalysisdiversemethodologies} to molecular unfolding \cite{bishwas2024molecularunfoldingformulationenhanced}. Furthermore, advancements such as parallel quantum annealing have shown promise in simultaneously optimizing multiple problems, highlighting the rapidly evolving capabilities of this technology \cite{bishwas2024investigationpotentialparallelquantum}.

The primary bottleneck in applying QA to large-scale logistics is the limited capacity and connectivity of current quantum hardware. Our analytical complexity comparison (Table \ref{Comparison}) demonstrated that Model B reduces the number of binary variables by over $50\%$ and rehandle-specific constraints by over $80\%$ compared to the conventional approach (Model A). This dimensional reduction is key, as it dramatically lessens the resources (qubits and auxiliary variables) required for a successful QUBO transformation (e.g., $\min_{x\in\{0,1\}^n} x^\top Q x$) and subsequent embedding onto quantum annealers. Since Model B relies entirely on binary variables and its objective function features bilinear products that map naturally to quadratic interactions, it provides a highly compatible foundation for QA. Standard linear constraints can be readily incorporated into the QUBO matrix $Q$ via penalty terms. While the exact QUBO formulation and hardware embedding are deferred to future work, this compact structure is a crucial first step toward scalable quantum logistics solutions.

\section{Conclusion}

This study successfully addresses the computational burden of the TLO problem. We found that the main source of complexity in existing models—the need to explicitly define every rehandle operation—could be eliminated by introducing a new, significantly simpler formulation that calculates rehandle costs directly within the objective function. Our core achievement is a substantial reduction in model size, with analytical comparisons showing that this approach cuts the number of variables by over $50\%$ and constraints by over $80\%$ in typical cases. This fundamental simplification is what makes the TLO problem far more manageable to solve.

The practical value of this compact model was confirmed through computational experiments using a simulated annealing metaheuristic. This approach efficiently generated high-quality loading plans for various scenarios, all within operationally acceptable time limits. These solutions are key for rail logistics, as they demonstrate how to maximize resource use while sharply reducing costly rehandle operations. However, our current model does have limitations, as it doesn't yet account for real-world complexities like container due dates, multi-train scheduling, or destination-specific loading blocks.

Moving forward, future work should focus on expanding the model to include these more complex operational realities. A valuable next step would be to rigorously test commercial integer programming solvers against our compact formulation and the developed metaheuristic, as well as compare our results with other methods like genetic algorithms or tabu search. 

\bibliographystyle{IEEEtran}
\bibliography{IEEEabrv,references}

\end{document}